# Revolutionizing student course selection: Exploring the application prospects and challenges of blockchain token voting technology

**Tiansu Hu[1,5,†], Yuzhao Song [2,†] Linjing Zhang[3,†], and Xiaoya Zhou[4,†]**

[1]Beijing-Dublin International College, Beijing University of Technology, Beijing, 100124, China

[2]Accounting Institute, Central University of Finance and Economics, Ningxia, 750000, China

[3]School of Medical information engineering, Jining Medical University, Qingdao, 266000, China

[4]Computer Science and Technology, Donghua University, Shanghai, 200051, China

[5]hutiansu@emails.bjut.edu.cn

[†]These authors contributed equally.

**Abstract.** This paper explores the utilization of blockchain token voting technology in student course selection systems. The current course selection systems face various issues, which can be mitigated through the implementation of blockchain technology. The advantages of blockchain technology, including consensus mechanisms and smart contracts, are discussed in detail. The token voting mechanism, encompassing concepts, token issuance and distribution, and voting rules and procedures, is also explained. The system design takes into account the system architecture, user roles and permissions, course information on the blockchain, student course selection voting process, and course selection result statistics and public display. The technology offers advantages such as transparency, fairness, data security and privacy protection, and system efficiency improvement. However, it also poses several challenges, such as technological and regulatory hurdles. The prospects for the application of blockchain token voting technology in student course selection systems and its potential impact on other fields are summarized. Overall, the utilization of blockchain token voting technology in student course selection systems holds promising future implications, which could revolutionize the education sector.

## 1. Introduction

The issue of course selection has always posed a challenge for both students and universities. During the concentrated course selection period, traditional systems frequently experience severe congestion due to increasing numbers of students and limited server processing capacity [1], which can negatively impact students' course experience.

Blockchain is a decentralized public ledger based on a peer-to-peer network [2]. Its decentralization, openness, and transparency can enhance the credibility and security of the system while also enabling distributed storage and management of course selection information, improving fault tolerance and scalability. Additionally, tokens are currency-like digital assets with payment and designated circulation functions [3, 4].

To integrate blockchain technology into the course selection system, we propose the introduction of a token voting system. Blockchain technology allows for the issuance of tokens, voting, and trading functions, which can facilitate students' course selection operations and result statistics. This system can improve the fairness and efficiency of course selection while also reducing pressure on centralized server resources. This paper discusses how to leverage blockchain technology and a token voting system to enhance university course selection systems. This approach offers a new perspective for addressing the challenges of course selection in academia.

## 2. Blockchain technology overview

This chapter will provide an overview of blockchain technology, focusing on its key components and their roles in supporting various applications. Blockchain technology adopts a chain-like structure, and its immutability and non-counterfeit ability characteristics are ensured through cryptography, combined in chronological order. This is a reliable technology for recording network change data transactions, with features such as security, decentralization, transparency, and immutability [5].

In the context of the student course selection system based on blockchain token voting technology, we will discuss how consensus mechanisms and smart contracts work together to create a secure and transparent environment for managing course selection data. Since 2020, the application of blockchain technology in the education field is still in its initial stage, mainly used for method verification and sharing certificates [6]. These practices have demonstrated that blockchain technology can provide a reliable and secure distributed environment, capable of carrying and recording data such as students' course selection.

### *2.1. Consensus mechanisms*

Consensus mechanisms are the process of nodes in a blockchain network reaches an agreement on a version of transaction. The security and reliability on blockchain network is ensured by consensus mechanisms, to make it difficult for attackers to tamper the recorded transactions. Ethereum currently uses a Proof of Stake (PoS) consensus mechanism. These are the main features of the PoS consensus mechanism:

a. Validator's staking: In the PoS mechanism, nodes, or validators, need to stake their tokens as collateral to obtain the rights of produce blocks and validate transactions. In Ethereum, a minimum of 32 ETH is required to become a validator.

b. Block Production and Validation: When new transaction occurs, validators compete for the right to produce the blocks. The validator who obtains the block production right will choose transactions to be packed in a block, create a new block, and adds it into the blockchain. After new block is created, other validators need to verify its correctness.

c. Rewards and Penalties: Validators earn rewards through staking and block production. If validators behave maliciously, including tamper with transactions, or trying to create invalid blocks, their staked tokens will be forfeited.

The Proof of Stake mechanism has higher energy efficiency and security compared to other consensus mechanisms, such as Proof of Work (PoW), another mechanism which Bitcoin applies.

### *2.2. Smart contracts*

Smart contracts are self-executing, blockchain-based contracts. They are written in specific programming language, to be executed on-chain. When certain predetermined conditions are met, smart contracts will execute automatically to do its operations. Ethereum is the first blockchain platform to support smart contracts, which are typically written in Solidity programming language. The main features of smart contracts are:

a. Decentralization: Smart contracts are deployed on blockchain networks; therefore, they are not subject to the control to any individuals. This enables higher transparency and immutability.

b. Automatic Execution: Smart contracts execute automatically according to preset conditions, eliminating the need for human involvement and reduce the likelihood of errors.

c. Programmability: Smart contracts like Ethereum smart contracts, are Turing-complete, which allows developers to create smart contracts with complex functionality. This has enabled the development of various decentralized applications (DApps) on Ethereum.

## 3. System design

This chapter will introduce the implementation process of a student course selection system based on token voting, which is deployed on a decentralized blockchain network that supports smart contract technology. This system aims to provide a safe, transparent, and fair course selection platform, enabling students to choose courses more conveniently and efficiently. The overall process of voting and course selection is controlled by smart contracts on the blockchain, and every interaction with the smart contract will be written into the block, making the course selection process trustworthy and tamper proof. The main implementation process of this system includes writing a smart contract for voting course selection schemes, designing a method for distributing tokens to students, and implementing a web front-end for course selection systems that interact with blockchain smart contracts.

*3.1. System architecture*

The student course selection system based on token voting is deployed on a decentralized blockchain network that supports smart contract technology. Ethereum is a decentralized blockchain network that has the characteristics of security, stability, transparency, and openness, and supports smart contract technology. Its characteristics can eliminate the need for centralized servers, so the Ethereum blockchain can carry the deployment of this system. The next system architecture will assume deployment on the Ethereum blockchain network.

The voting protocol, including the management and execution of the entire voting process, is implemented on the Ethereum smart contract. The smart contract will automatically run on the Ethereum chain after being initialized by the voting initiator. During operation, the smart contract will control the voting process and automatically enter the next stage at the predetermined time.

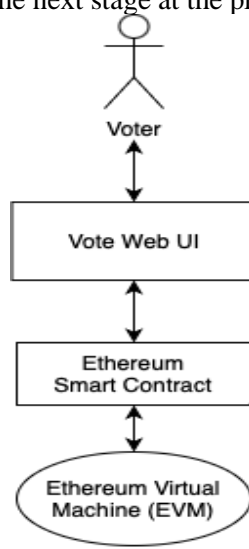

**Figure 1.** schematic diagram.

The schematic diagram of this project is shown in Figure 1, where each voter's voting process is as follows: a vote is initiated to the selected course through the front end of the webpage, the webpage interacts with the smart contract, a transaction is initiated to the smart contract address, and the corresponding method in the smart contract is activated. When the operation result of the smart contract on the blockchain is reported as successful, the user successfully participates in the voting. The smart contract is deployed on the Ethereum blockchain, and its compiled code is distributed and stored in each node of the blockchain network. When the smart contract is activated, it is executed in a decentralized manner by the Ethereum Virtual Machine (EVM), ensuring that each node on the network has a copy of the data, thereby ensuring the stability of the decentralized voting and course selection system.

The abilities that students (voters) possess include determining the courses they vote for, as well as determining the number of votes to be cast for the selected courses; The Vote Web UI is a page that interacts with students and has the ability to display and interact. In terms of display, the front end of the webpage will display course information, remaining tokens, voting time, and the running results of course selection requests, etc., to facilitate students' voting decisions. In terms of interaction, the front-end of the webpage will obtain the course ID and number of votes selected by the user, and then initiate a request to the Ethereum smart contract. Ethereum Smart Contract is the core component of the entire system, which has the following capabilities: controlling, generating transactions, and generating results. In terms of control, the smart contract controls the voting stage, where self preparation, voting, and statistical results are conducted in a one-way stable manner, and the legitimacy of the voting is verified (time, number of tokens, voters, etc.). In terms of generating transactions, smart contracts will record the results of each vote (success/failure) and record them on the blockchain. In terms of generating results, after the voting is completed, running the corresponding method on the smart contract can return the final statistical results of course selection, which is the final winner of each course.

*3.2. Voting rules and mechanisms*
In this course selection system, voting rules and mechanisms are implemented through smart contracts, aiming to ensure security, stability, and transparency, so that each student can choose their own satisfactory course combination according to their own needs as much as possible.

The specific details of the implementation of this system are as follows: firstly, the smart contract will allocate optional courses for the voting stage based on the number of credits taken by students in the semester determined during the declaration stage. After entering the voting stage, students are free to invest their tokens into the required courses. After the voting phase ends, enter the settlement phase. The settlement phase will be processed by smart contracts, with a specific allocation algorithm that sorts each course based on the number of tokens put in by students. Then, the top n students ranked in the course capacity are selected as successful candidates for the course. From this, it can be seen that students who invest more coins have an advantage in the competition, making them more likely to obtain the courses they expect.

If a student fails to successfully choose a certain course, they still have the opportunity to secure a spot in other less competitive courses. For this group of students, the system will process other courses one by one in order of their vote count, from highest to lowest. The processing of other courses will follow the principles of no time conflict and no repetition of courses. In the courses that meet the requirements for course selection, the system will arrange corresponding courses for these students until the pre agreed number of credits is selected.

*3.3. Token issuance and distribution*
The school will determine and publish the total academic scores that students in each major should take based on the teaching plan and teaching situation, for reference by students in different majors. In the token voting course selection system, before the course selection stage of each semester, students need to declare to the smart contract and set the number of course credits they decide to take [7]. The smart contract will distribute a corresponding number of tokens to different students in a proportional proportion to the required credits according to their needs. The tokens for each semester are independent and do not exist for inheritance between semesters. For example, the "Spring2023" Token released for course selection in the spring of 2023 can only be used for course selection in the spring semester of 2023. In the summer semester of 2023, a new 'Fall2023' token will be released, while the outdated

'Spring2023' token will become invalid and no longer be used for voting. The setting of this mechanism helps to ensure that the average token cost for all courses is the same.

In the course selection stage, each student needs to refer to different course time periods and make preliminary plans for course selection. And achieve a targeted and arranged allocation of tokens, ultimately throwing a reasonable amount of tokens for each course that the student believes is reasonable. In each semester's course selection, students need to find ways to ensure that all types of elective courses meet the academic scores set by the college, in order to meet the requirements for smooth graduation.

## 4. System analysis and discussion

*4.1. Comparison between token voting system and traditional course selection system*
The traditional course selection system requires high server response time, and some students may miss out on courses due to external factors such as internet speed. The token voting system refers to students throwing tokens at their desired courses based on their own planning, overall arrangement, and careful consideration. In comparison, the token voting system can better promote students' thinking on resource allocation, guide them to choose courses that are in line with their professional plans, and guide them to explore the future path [8]. Moreover, the token voting system for course selection to some extent avoids the drawbacks of traditional course selection systems, which result in server paralysis caused by a large number of students entering the system together in a short period of time. However, in terms of system design, the token course selection system requires higher openness, computability, and sorting of data, so it has higher requirements for algorithms and compatibility.

*4.2. Evaluation and limitation analysis of token voting system in course selection system*
During peak periods of high resource demand, traditional network architectures have poor global performance and uneven distribution of real-time traffic, resulting in poor network timeliness. The token voting system eliminates the need for centralized servers and slows down the surge in access to course selection systems due to limited quotas during centralized time periods, resulting in excessive thread blocking, low server information processing rate, extremely slow response speed, and severe congestion and lag [9]. This system can meet the course selection needs of most students in a short period of time, reduce the time cost and mental pressure of course selection, and greatly improve the course selection experience. At the same time, this mechanism to some extent reduces the consumption of server resources and reduces the maintenance costs of the school.

The introduction of token voting system in the course selection process has strong innovation and also solves the practical difficulties faced by students. Before each semester's course selection, students determine and declare to the system the number of courses and corresponding credits they have decided to choose for this semester. The system also maximizes the satisfaction of students' initial wishes during course selection based on their wishes. This not only gives students the right to make independent choices, but also allows them to arrange their four-year university courses according to their own development needs. And to some extent, it tests students' ability to coordinate and plan as a whole. The amount of credit applications and coins invested in each semester is important, reminding students to carefully consider every choice they make.

The token voting system also has limitations. Firstly, this system carries certain investment risks and cannot prevent a small number of students from spending too much tokens in the early stage due to gambler mentality or poor arrangements. As graduation approaches, there are still many courses that are not attended due to insufficient tokens, and ultimately they fail to complete the required total academic scores and credits for various types of courses in their respective majors, resulting in unsuccessful graduation. Secondly, due to the introduction of the token system, the essence of course selection has shifted from the traditional "first come, first served" approach of seizing network resources to a "multi coin" bidding approach. Therefore, the competition process cannot guarantee complete fairness, and there will be token depletion for some high investment bidders [10].

## 5. Conclusion

The application of the token voting system in course selection processes has significant implications. Not only does it improve the student course selection experience and optimize the course selection system, but it also has the potential to align closely with national conditions and create added value. Through this system, the popularity of different courses and various content among students can be intuitively reflected, serving as an evaluation criterion for assessing teachers' teaching abilities. In the future, blockchain technology is expected to enable information sharing between teachers and students nationwide or globally. By comparing the number of tokens allocated by students to different courses or teachers of the same course and collecting corresponding data through real-time questionnaires, a comprehensive analysis can be conducted. This allows teachers to always know students' needs and optimize course arrangements to improve students' course selection experiences. Through semester course openings, students using token course selection, questionnaire recycling, and data statistical analysis can achieve a perfect closed loop of mutual understanding between students and teachers, realizing a perfect two-way communication mode.